# Astro2020 APC White Paper

**Title: A Great Successor to the Hubble Space Telescope**
**Type of Activity: Space Based Project**
**Principal Author:**

Name: B. Scott Gaudi
Institution: The Ohio State University
Email: gaudi.1@osu.edu
Phone: 614-292-1914

**Co-authors:** John C. Clarke, Boston University, jclarke@bu.edu, Shawn Domagal-Goldman, GSFC, shawn.goldman@nasa.gov, Debra Fischer, Yale University, debra.fischer@yale.edu, Alina Kiessling, JPL, Alina.A.Kiessling@jpl.nasa.gov, Bertrand Mennesson, JPL, Bertrand.Mennesson@jpl.nasa.gov, Bradley M. Peterson, Ohio State University/STScI, peterson.12@osu.edu, Aki Roberge, GSFC, Aki.Roberge@nasa.gov, Dan Stern, JPL, Daniel.K.Stern@jpl.nasa.gov, Keith Warfield, JPL, keith.r.warfield@jpl.nasa.gov

**Description**:

The Hubble Space Telescope (HST) has been the most impactful science-driven mission ever flown by NASA. However, when HST eventually reaches the end of its life, there will be a void due to the loss of some of the science capabilities afforded by HST to astronomers world-wide. In particular, no other existing or planned observatory can undertake high-resolution UV imaging and spectroscopy. The previous 2010 Decadal Survey (DS) noted this void, arguing for the need for a successor to HST with UV capabilities in three separate places in the main report (pp. 190, 203, and 220). They further noted that realizing such a mission called for further technology development, specifically detectors, coatings, and optics. The large strategic missions that will follow HST, namely JWST and WFIRST, will continue to spark the interest of the public in space-based astronomy. However, in order to ensure continued US preeminence in the arena of large strategic space-based astrophysics missions, as well a seamless transition after WFIRST, a future flagship mission must be "waiting in the wings." Anticipating this need, NASA initiated four candidate large strategic mission concept studies (HabEx, LUVOIR, Lynx, and Origins), which have advanced, mature designs, including detailed technology assessments and development plans. Two of these concepts, HabEx and LUVOIR, are responsive to the recommendations of the previous Decadal Survey regarding a UV-capable mission. Either represent a more powerful successor to HST, with UV-to-optical capabilities that range from significant enhancements to orders-of-magnitude improvement. At the same time, technological and scientific advances over the past decade only now make it feasible to marry such an astrophysics mission with one that can search for life outside the solar system. Acknowledging that the constraints that the Astro2020 DS must consider may be difficult to anticipate, the HabEx and LUVOIR studies together present eleven different variants, each of which enable groundbreaking science, including the direct imaging and characterization of exoplanets. The HabEx and LUVOIR mission studies therefore offer a full suite of options to the Astro2020 DS, with corresponding flexibility in budgeting and phasing.

*The Legacy of the Hubble Space Telescope (HST)*

There is perhaps no more iconic and immediately recognizable image of a space telescope than that of HST (**Figure 1**). HST was launched on April 24, 1990, and is approaching its 30th anniversary. HST has inspired generations of scientists, with such iconic images at the "Pillars of Creation", the "Hubble eXtreme Deep Field", and extraordinary images of resolved stellar populations (**Figure 2**).

HST has made important contributions to nearly every area of astronomy and planetary science, resulting in over 16,000 refereed publications. While these contributions are too numerous to list here, it is worth highlighting a few notable examples uniquely enabled by space-based observations at wavelengths shorter than 600nm. We will also highlight how improved UV/Optical capabilities, such as will be available with HabEx or LUVOIR, would substantially or dramatically improve these science cases.

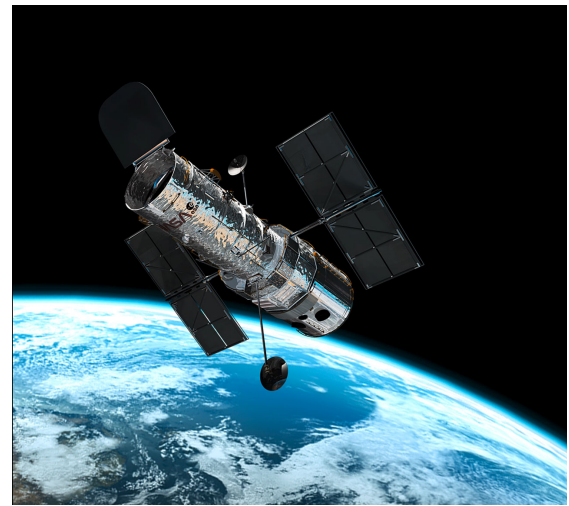

**Figure 1.** The Hubble Space Telescope from orbit. NASA/ESA

**Measuring the Hubble Constant**

One of HST's original three key projects was to measure the Hubble-Lemaître Constant ($H_0$), to 10%. While this project used a variety of distance indicators to measure $H_0$ (e.g., Tully-Fisher, Type 1a SNe, surface brightness fluctuations, the fundamental plane, and Type II SNe), HST's contributions were essential. The high spatial resolution of HST allowed the detection of Cepheids (and thus Cepheid period-luminosity relations) out to distances of roughly ten times that could be done from the ground. This allowed Cepheid calibration of these secondary

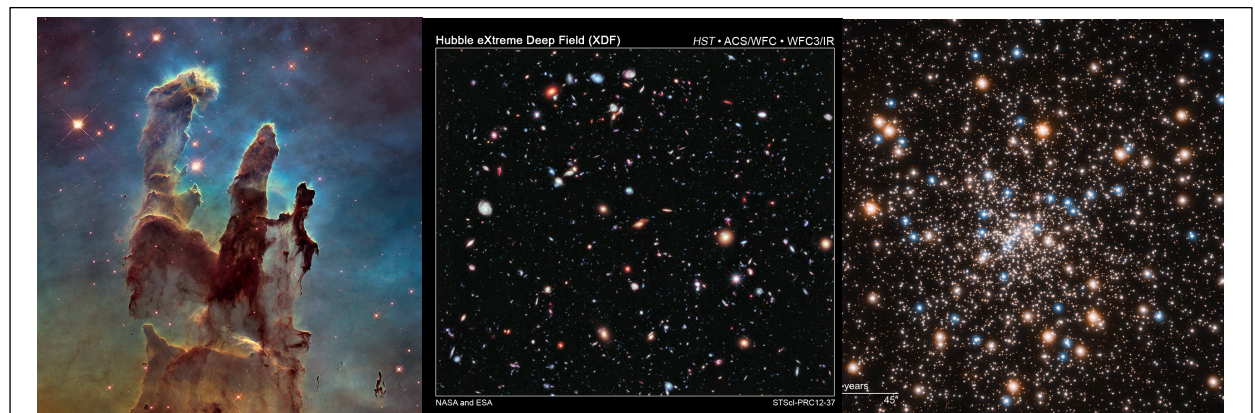

**Figure 2.** (*Left*) The "Pillars of Creation", a WFC3 image of the Eagle Nebula, and one of the most iconic images ever taken by HST. (*Middle*) The Hubble eXtreme Deep Field (XDF), one of the deepest images of the sky ever taken, with a total exposure time of roughly 2 million seconds. (*Right*) The Globular Cluster NGC 6397, whose distance was measured with HST using a new spatial scanning mode. Credit: NASA

indicators, most crucially for Type 1a SNe, which previous to HST had no Cepheid calibrators (Freedman et al. 2001).

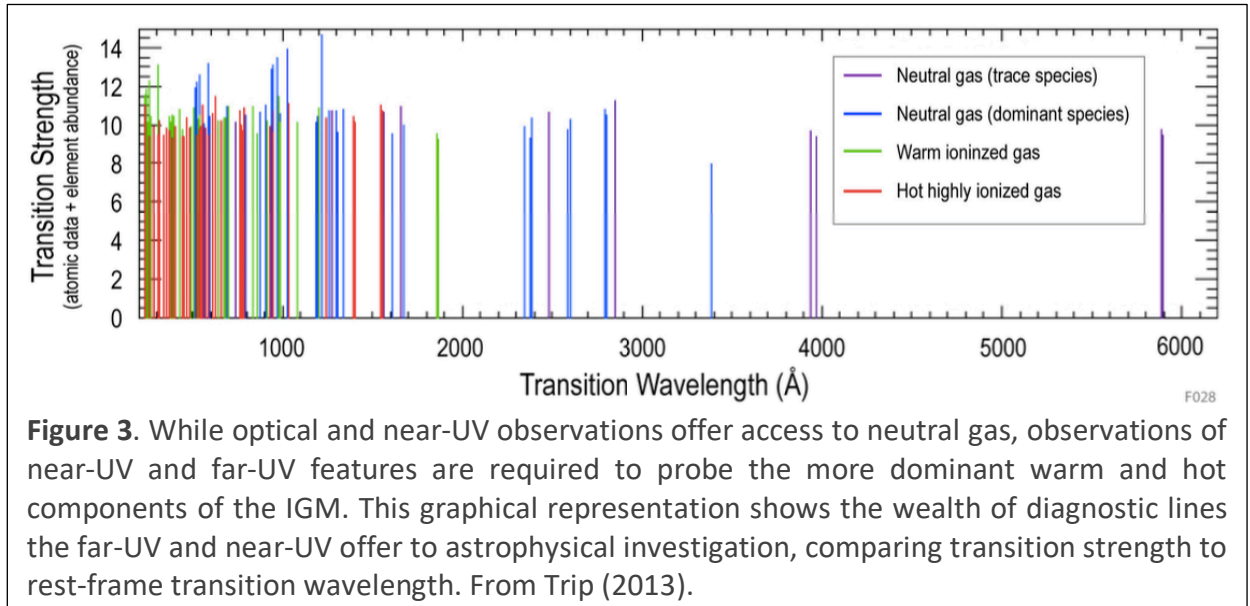

**Figure 3**. While optical and near-UV observations offer access to neutral gas, observations of near-UV and far-UV features are required to probe the more dominant warm and hot components of the IGM. This graphical representation shows the wealth of diagnostic lines the far-UV and near-UV offer to astrophysical investigation, comparing transition strength to rest-frame transition wavelength. From Trip (2013).

Freedman et al. (2001) were indeed successful in measuring $H_0$ to (nearly) 10%, finding $H_0$=72+/-8 km s$^{-1}$ Mpc$^{-1}$. Nevertheless, the SH0ES Team (Riess et al. 2016) picked up the gauntlet in an effort to improve the precision of the measurement of $H_0$ with newly-available observing methods and better control of systematics. One particularly notable novelty in the approach of the SH0ES team was to use drift scanning to improve the precision of the parallax distances to Galactic Cepheids, as well as the relatively new Drift and Shift (DASH) observing mode, whereby new targets are imaged without the need to acquire guide stars, which results in higher efficiency. This allowed Riess et al. (2019) to monitor Cepheids in the LMC more efficiently. When combined with a ~1% uncertainty to the distance to the LMC by Pietrzynski et al. (2019), Riess et al. (2019) find $H_0$=74.03+/-1.42 km s$^{-1}$ Mpc$^{-1}$, roughly 4.4σ larger than the value inferred from Planck by the CMB and ΛCDM (The Planck Collaboration 2018).

The increased spatial resolution and sensitivity of both HabEx and LUVOIR would enable observations of Cepheids over a much greater distance, as far as the Coma Cluster with LUVOIR-A, thereby improving upon local distance ladder measurements. The increase in the number of Hubble flow galaxies with observations of both Cepheids and Type Ia SNe will greatly reduce uncertainty in the total error budget.

**The Life Cycle of Baryons**

Despite decades of efforts, approximately one-third of the baryons in the local universe remain unaccounted for. The "missing baryons" are thought to be predominantly in the form of diffuse, hot gas around and between galaxies, but many fundamental questions remain open about this gas, even within the very local universe. This material, the intergalactic medium (IGM; i.e., the gas between galaxies) and the circumgalactic medium (CGM; i.e., the gas external to, but near galaxies), is the fuel from which stars ultimately form, and, later in their lives, the material that galaxies redistribute and enhance through supernovae and violent mergers. Inflows and outflows of the CGM are inextricably linked to key issues, such as star formation, galactic structure, and galactic morphological transformation. Therefore, studying and understanding this gas is essential for understanding the life cycle of baryons in the cosmos, and for developing a more complete picture of galaxy formation and evolution. However, this presents observational challenges since the bulk (60%) of the CGM is predicted to be extremely hot, with the key diagnostic transitions at UV and X-ray energies and thus inaccessible to the ground (**Figure 3**). Thus, empirical studies of the CGM have relied primarily on its absorption signatures in the rest-frame UV spectra of bright background quasars. Observations to date, largely based on statistical studies built out of samples of single sightlines per vast galaxy halo (Tumlinson et al. 2013), show that the CGM is significantly metal-enriched, and that it is dynamic and short-lived. The results, largely based on observations by the Cosmic Origins Spectrograph (COS) on HST, show that metal-enriched, under-pressurized 'clouds' at galactocentric distances greater than 75 kpc appear to have no means of long-term survival, yet are commonly found in statistical studies of quasar absorption lines. In addition, vast reservoirs of neutral hydrogen surround both star-forming and passive galaxies alike, hinting that the lack of a gas supply cannot entirely explain the low levels of star formation in passive galaxies.

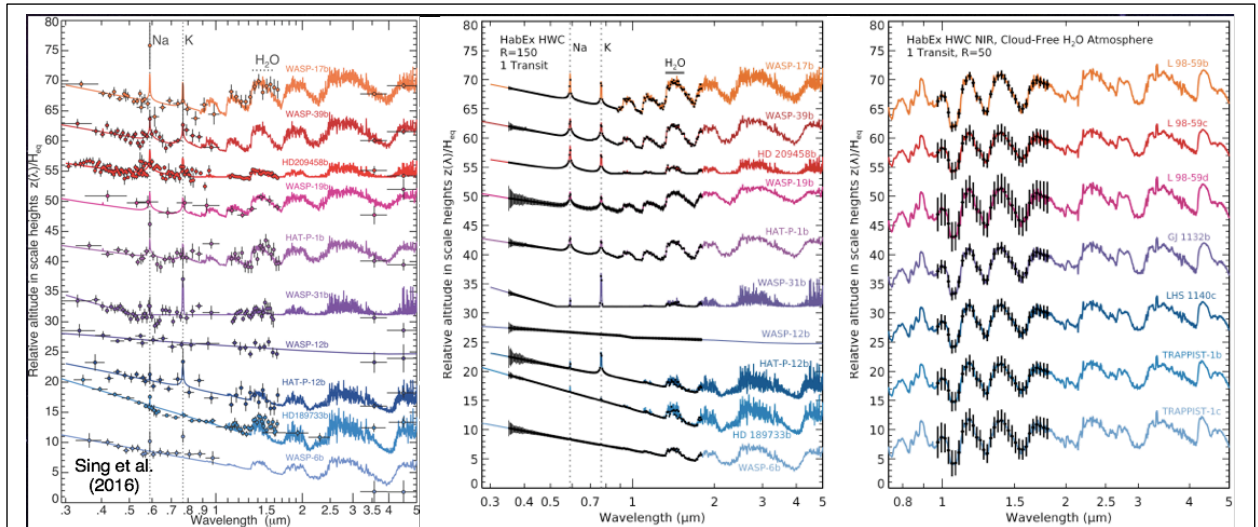

**Figure 4.** (*Left*) From Sing et al. 2016. HST/Spitzer transmission spectral sequence of hot-Jupiter survey targets. Solid colored lines show fitted atmospheric models with prominent spectral features indicated. (*Middle*) Simulations of the transmission spectra on the left panel, as would be measured by HabEx. The uncertainties are sufficiently small that simulations of observations by LUVOIR of these spectra would be indistinguishable by eye. (*Right*) Simulated spectra of the spectra of rocky planets transiting M dwarfs observable by HabEx. The uncertainties for LUVOIR would be ~2.3 times smaller (LUVOIR B) and ~4.8 times smaller (LUVOIR A). Note that while HabEx cuts off at 1.8 μm, LUVOIR's wavelength range extends 2.5 μm. Middle and Right panels simulated by E. Lopez (GSFC).

To make significant progress in constraining and understanding the cosmic baryon cycle, it is necessary (1) to complete the census of baryons in the local universe; (2) to measure the amount of gas and heavy elements around z < 1 galaxies; and (3) to determine the dynamical state and origin of the various components of the IGM, i.e., determine what fraction of the IGM is primordial, due to outflowing material, recycled accretion, or other physical causes.

Both HabEx and LUVOIR would improve our understanding of structure and physics of the CGM by increasing the number of sightlines per halo of background quasars, which can be used as tracers of the CGM through the absorption lines in the quasar spectra. Here the advantage is twofold. Fainter quasars will be accessible due both to the larger collecting area and improved throughput, and the observations can be multiplexed through the use of microshutter or micromirror arrays.

**Characterizing the Atmospheres of Exoplanets**

HST was not designed to study the properties of exoplanets. Nevertheless, it has been transformational in our understanding of exoplanets, particular transiting exoplanets. Brown et al. (2001) ushered in the age of studying transiting planets with space-based facilities by using the STIS spectrograph to achieve extremely precise relative photometry of HD209458, resulting in tight constraints on the properties of the star and the transiting planet, and ruled out the existence of large satellites or rings. Later, Charbonneau et al. (2002) used the same dataset to provide the first detection of the atmosphere of an exoplanet via transmission spectroscopy.

Perhaps the highest precision achieved with HST has been through the spatial scanning mode. This was used by Kriedberg et al. (2014) to obtain a spectrum of the Super-Earth GJ1214 with a precision of better than 50 ppm, enabling them to rule out high mean molecular weight atmosphere, and thus conclude that low-pressure clouds were in abundance in the atmosphere or GJ1214.

Both HabEx and LUVOIR will have the high-resolution sensitivity to study the atmospheres of transiting planets in the key region shortward of 600 nm, extending to at least 120 nm in the FUV. This will allow them to measure atmospheric escape from a wider range of warm-to-hot exoplanets than is possible with Hubble (e.g., Ehrenreich et al. 2015). This capability also permits detection of Raleigh scattering (an important indicator of atmospheric pressure) and ozone (an important biosignature for oxygen-poor, inhabited planets like the Proterozoic Earth). Thus, HabEx and LUVOIR will dramatically increase our inventory of the atmospheres of transiting planets, and in particular will be sensitive to wavelengths inaccessible to JWST or from the ground **(Figure 4)**

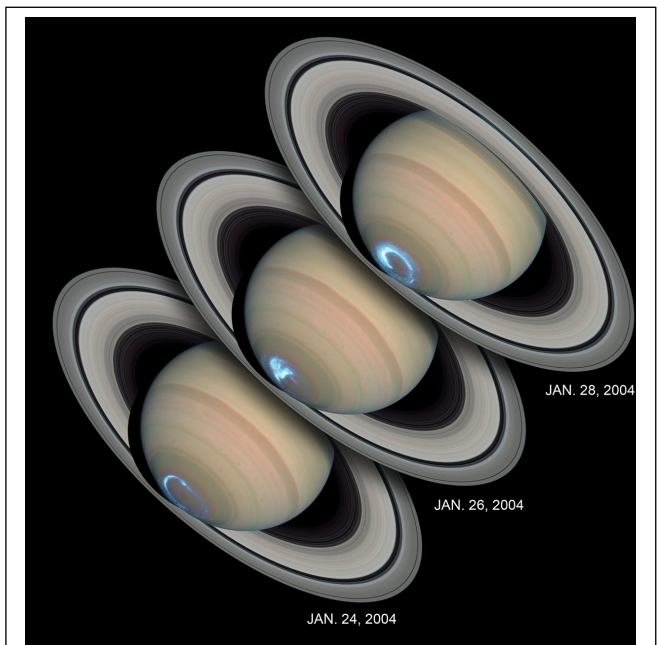

**Figure 5**. HST images that reveal the dynamic nature of Saturn's aurorae. The images combined ultraviolet observations taken HST's STIS instrument with visible images were taken with the ACS instrument. **Credit: NASA/ESA,** John Clarke (BU) and Z. Levay (STScI)

**Aurorae of the Solar System Giant Planets**

Planetary aurorae are visible examples of star-planet interactions. Aurorae are the manifestation of planetary magnetic fields interacting with stellar winds, and dictate the extent to which planetary auroral activity is driven by stellar winds as opposed to plasma processes in the planetary magnetosphere. Aurorae have also been seen on all of the gas giants of our solar system, as well as some giant-planet moons (e.g., Figure 3). Solar system aurorae cover a wide range of physical scales, conditions, and timescales, thereby providing an important testing ground for probing star-planet interactions in exoplanetary systems. For example, what controls auroral processes on different scales of time and planet size, different levels of stellar winds, different planetary rotation rates, and different magnetic field strengths? On the Earth, the solar wind's flow time past the planet is a few minutes, and auroral

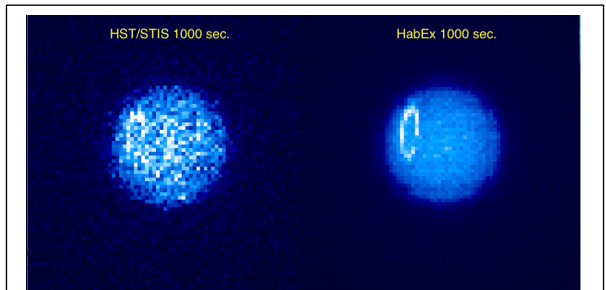

**Figure 6.** *(Left)* Simulated images of Uranus from HST. *(Right)* Simulated images from HabEx. In both panels, the exposure time is 1000s. In the right panel, the aurora is clearly visible. LUVOIR would have both higher resolution and higher sensitivity. **Credit:** John Clarke (BU)

storms develop in a complex interaction with the Earth's magnetic field in conjunction with the interplanetary magnetic field. On Jupiter and Saturn, the flow time is hours to days. Jupiter sometimes responds to changes in the solar wind, other times not at all, while Saturn's auroral activity seems to respond to every solar wind pressure front. Is auroral activity at Saturn controlled just by solar wind pressure, or is the interplanetary magnetic field direction important? An open question is whether Saturn's aurora is similar to the Earth's, or whether it has a different interaction with the solar wind.

Extended high-resolution investigations to Neptune, and improving the observational capability beyond that of HST at distances of Uranus and greater, will enable access to different configurations of internal magnetic fields that are highly tilted and offset from the planets' rotation axes. This would provide solar system analogs to the large number of 2–5 Earth-radii exoplanets recently discovered by *Kepler*. **Figure 6** shows a simulation of the ability of a UV-optimized telescope with an aperture as small as the preferred architecture of HabEx to image the aurorae on Uranus. It is noteworthy that the larger effective area not only allows HabEx to collect photons at a higher rate, but enables higher time resolution in the shape and brightness of the aurorae, which cannot be achieved by simply integrating longer.

**The relevance of wavelength coverage in the UV and visible**

We note that all four science applications discussed above utilized observations at wavelengths shortward of the cutoff of JWST and WFIRST (~600nm). For example, both Freedman et al. (2001) and Riess et al. (2019) used observations in the F555W (V band, 5550Å) to discover Cepheids to calibrate the period-luminosity relationships used to calibrate secondary distance indicators and thus measure $H_0$. Observations of planetary aurorae were made by HST's STIS instrument. UV observations of transiting planets are critical for detecting Raleigh scattering or the presence of high-altitude clouds or hazes. Simply put, astronomical observations can, of course, make progress in any single waveband, but history has shown that multi-messenger astronomical capability's is critical, simply because astronomical phenomenon are inherently panchromatic.

## We ARE NOT Running Out of Things to Do With HST.

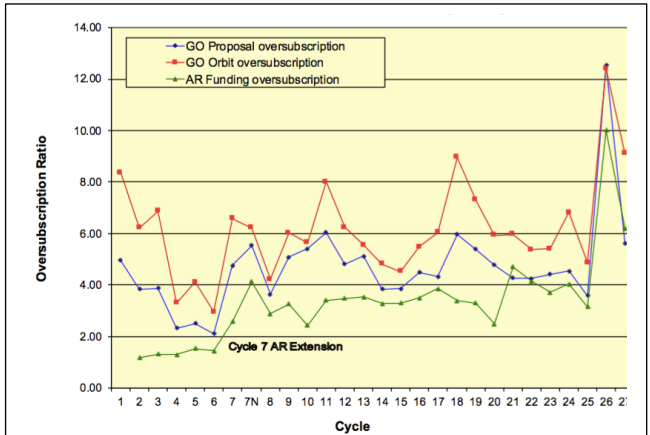

**Figure 7**. The over-subscription rate for HST by cycle. Credit: STScI

There is no indication that HST is reaching the end of its scientifically productive lifetime, even after nearly 30 years (we note that this does not imply that we don't require improved capabilities relative to HST for some science applications). **Figure 7** shows the HST oversubscription rates as a function of cycle, including both GO and AR proposals. Generally, the panel over-subscription rates have remained at an average of ~6 per cycle, with significant fluctuations.

## The Need for a Successor to HST with Ultraviolet Capabilities

There have been five HST servicing missions flown by the space shuttle; the most recent just over a decade ago in May 2009. HST has remained fully functional since the last mission. Hubble is expected to continue science operations through at least 2025. **Figure 8** shows that the reliability of all components (including instruments and subsystems) are predicted to be >80% through 2025, with the exception of the 3 gyros. Nevertheless, it is expected that there will be at least one gyro well past 2025; only one gyro is need for operations.

However, HST will eventually reach the end of its life, and when this happens, there will be a void due to the loss of some of the science capabilities afforded by HST to astronomers world-wide. In particular, no other existing or planned observatory can undertake high-resolution UV imaging and spectroscopy. The previous Astro2010 DS, "New Worlds, New Horizons (NWNH)," noted this void, stating in particular that:

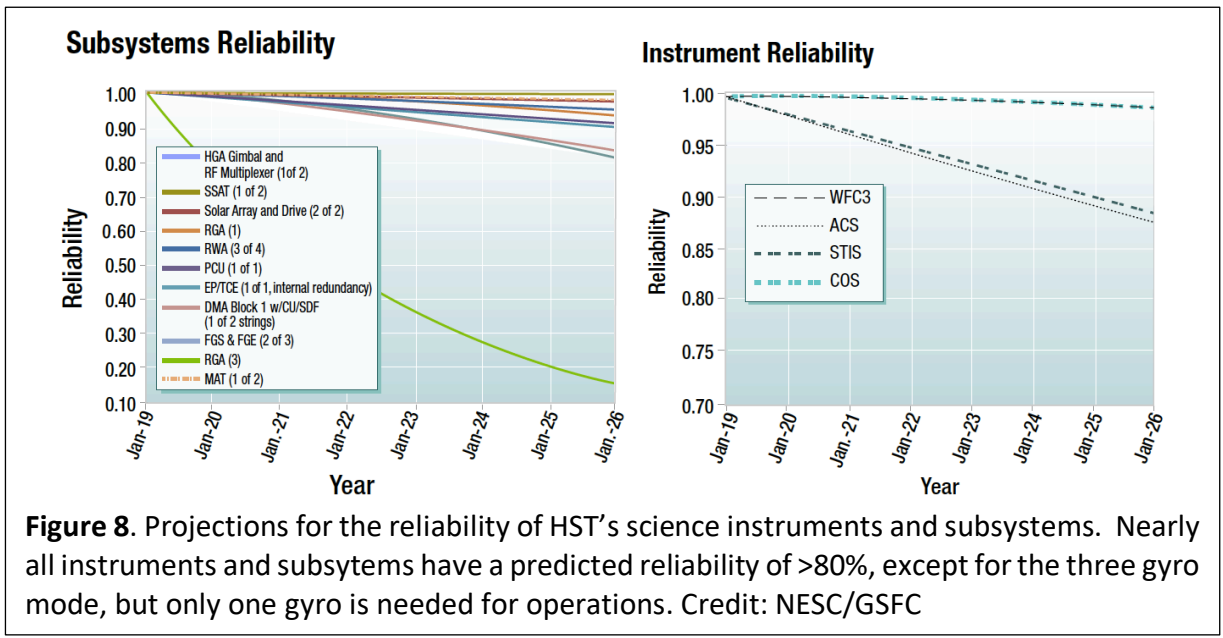

**Figure 8**. Projections for the reliability of HST's science instruments and subsystems. Nearly all instruments and subsytems have a predicted reliability of >80%, except for the three gyro mode, but only one gyro is needed for operations. Credit: NESC/GSFC

"Studies of the intergalactic medium, which accounts for most of the baryons in the universe, at more recent times could be transformed by an advanced UV-optical space telescope to succeed the Hubble Space Telescope (HST), equipped with a high-resolution UV spectrograph." – p. 190

"The cycling of gas from galaxies to the surrounding intergalactic medium and back again could also be studied with a ground-based Giant Segmented Mirror Telescope, using high-resolution optical spectra to study gas absorption lines highlighted by background quasars along many sight-lines, but a future UV space mission will be needed for a complete inventory." – p. 203

"Key advances could be made with a telescope with a 4-meter-diameter aperture with large field-of-view and fitted with high-efficiency UV and optical cameras/ spectrographs operating at shorter wavelengths than HST." –p. 220

They further noted that realizing such a mission called for further technology development, specifically with regard to detectors, coatings, and optics.

### *The Importance of Large Strategic Space-Based Missions and Maintaining Continuity*

The large strategic missions that are the successors to HST, namely JWST and WFIRST, will continue to spark the interest of the public in space-based astronomy. However, in order to ensure continued US preeminence in the arena of large strategic space-based astrophysics missions, as well as to ensure a seamless transition after WFIRST, a future flagship mission must be "waiting in the wings." Moreover, identification of a future large mission at least five years in advance of Phase A enables focused technology development that will result in long-term savings. Anticipating this need, NASA has gone to great lengths to ensure that there exist four candidate large strategic mission concepts, specifically HabEx, LUVOIR, Lynx, and Origins, that have advanced and relatively mature designs, as well as detailed technology assessments and development plans.

### *The Four Large Mission Concepts*

On January 4, 2015, NASA Astrophysics Division (APD) Director Paul Hertz charged the astronomical community – specifically the three APD Program Analysis Groups (COPAG, ExoPAG, and PhysPAG), to solicit community input to develop a short list of candidate large strategic mission concepts, which would then be studied in detail through mission concept studies conducted by NASA. In this white paper, Director Hertz suggested suggest four concepts for study: the Far-IR Surveyor, the UV/Optical/IR Surveyor, the X-ray Surveyor, and the Habitable Exoplanet Imaging Mission. After nearly 10 months of discussions, the PAGs concurred that all four large mission of these mission concepts should be candidates for mission concept maturation prior to the 2020 Decadal Survey. Other mission concepts were considered, as well as the idea of merging the UV/Optical/IR Surveyor with the Habitable Exoplanet Imaging Mission, but none achieved broad community support. These four mission concepts were subsequently renamed Origins, the Large UVOIR Surveyor (LUVOIR), Lynx, and the Habitable Exoplanet Observatory, respectively.

Work on these mission studies began in early 2016. Each study was assigned to a NASA center, and Science and Technology Definition Teams (STDTs) were assembled, each with two co-community chairs. The ultimate succinct goal of these STDTs was to issue a final report that includes a science case with proposed science objectives, a strawman payload, a design reference mission, and technology development required to enable a new mission start. The final reports are due on August 22, 2019.

The study teams, drawn from the broad scientific community and NASA, have worked for over three years alongside partners in industry and representatives of the international science community. Each team has spent many thousands of person-hours and millions of dollars to create the scientific and technological visions for their missions. As a result, we believe that these concepts are at a level of detail and rigor that is rarely seen for NASA missions at this early stage.

*Comparing the Primary Science Drivers and Capabilties of HabEx and LUVOIR*

While differing greatly in design and scope, the HabEx and LUVOIR mission studies have broadly similar primary general astrophysics science drivers. Both are responsive to the recommendation of the previous Astro2010 DS regarding the need for a UV-capable mission. Each represent a more powerful successor to Hubble, with UV-to-optical capabilities that range from dramatically enhanced to orders-of-magnitude improvements. This is not to say that both mission concepts have similar capabilities. Although some of the science capabilities scale smoothly with aperture size, in several cases there are clear break points below which some science becomes impossible or impractical. These break points are laid out in both final reports.

At the same time, technological and science advances over the past decade makes it not only feasible to marry such a mission with one that will also address one of the most profound questions of humankind: is there life outside the solar system? Although this is not a new question, we argue that it is one for which we can provide a meaningful answer.

Both HabEx and LUVOIR are Great Observatories capable of addressing some of the most fundamental questions in astrophysics and planetary science, as well as being able to directly detect and obtain spectra of Earth analogs, and thereby search for signs of habitability and perhaps even biosignatures. The LUVOIR A and B concepts will yield a relatively large sample of ExoEarth candidates, enabling a high-confidence constraint on the frequency of potentially habitable worlds with biosignatures. HabEx, on the other hand, while having a very low probability (<1.4%) of detecting no potentially habitable worlds, will obviously have a smaller sample size. However, as we emphasize below, the yield is not a simple function of aperture.

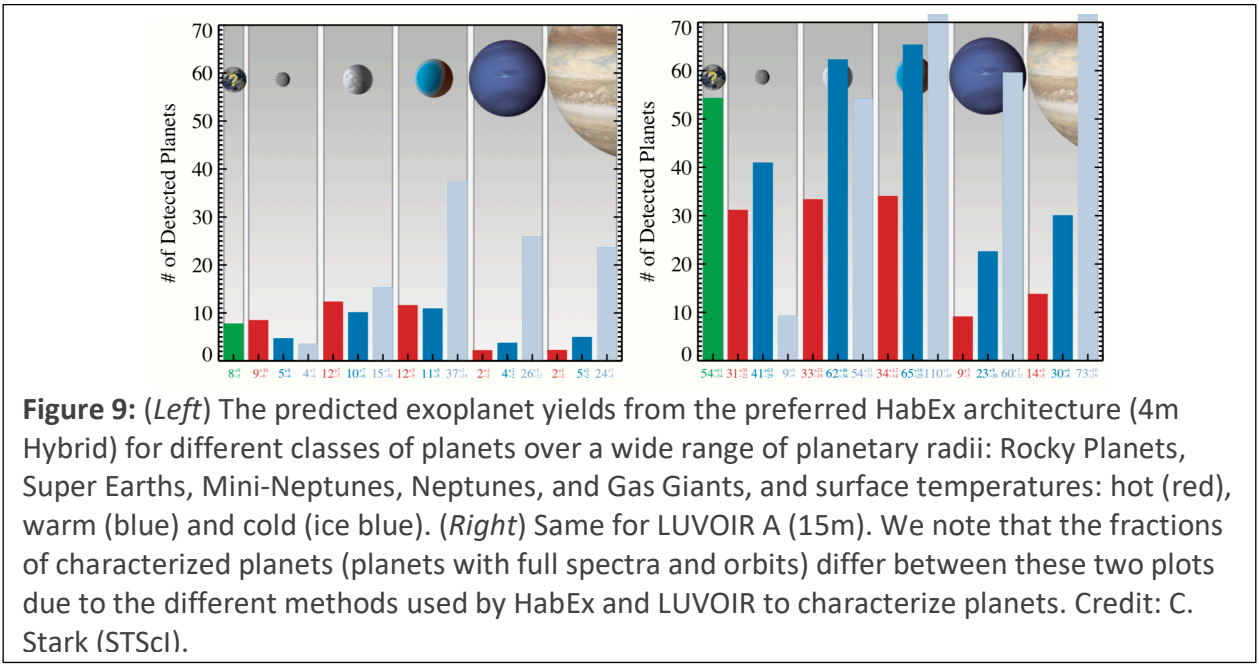

**Figure 9:** (*Left*) The predicted exoplanet yields from the preferred HabEx architecture (4m Hybrid) for different classes of planets over a wide range of planetary radii: Rocky Planets, Super Earths, Mini-Neptunes, Neptunes, and Gas Giants, and surface temperatures: hot (red), warm (blue) and cold (ice blue). (*Right*) Same for LUVOIR A (15m). We note that the fractions of characterized planets (planets with full spectra and orbits) differ between these two plots due to the different methods used by HabEx and LUVOIR to characterize planets. Credit: C. Stark (STScI).

*Expected Capabilites and Flexibilty to External Constraints*

The detailed capabilities of both of LUVOIR's architectures and of HabEx's preferred architecture will be presented in their corresponding white papers and final reports, which we will not reiterate here. However, we will summarize the prospects for detecting and characterizing mature exoplanets in reflected light, particularly ExoEarth Candidates to search for habitable conditions and biomarkers, as these are novel capabilities of HabEx and LUVOIR that were not only not available on HST, but largely drove the design of these missions (**Figure 9**).

As described in detail by Stark et al. (2019), designing optimal direct image mission architectures involves a number of trades, many of which can be non-intuitive and lead to poor yields despite large apertures. Both the HabEx and LUVOIR studies studied these trades in great detail and feel that they have arrived at "locally" optimal architectures and observing strategies for the particular aperture being considered.

Acknowledging that the constraints that must be considered by the Astro2020 DS may be difficult to anticipate, or may change over time, the HabEx and LUVOIR studies together present eleven different architectures. HabEx considered nine different architectures, and LUVOIR considered two architectures. All architectures offer an increase in the effective area in the UV and FUV limit over HST. Also, all architectures can directly image and characterize exoplanets, although for the smallest apertures considered by HabEx, not true Earth analogs.

The primary goal of this white paper is to emphasize that HabEx and LUVOIR present a suite of mission concepts that can achieve a broad range of exciting science that will replace and enhance that lost by HST, but will also include the capability to detect and characterize potentially habitable planets. These studies have been performed in sufficient detail and breadth to allow for prioritization of the science themes enabled by the HabEx/LUVOIR family, without prioritizing a specific point design. Indeed, we do not expect the DS to select one of the 11 specific designs presented by the HabEx and LUVOIR studies, but rather to consider the full suite as a continuum of options. They provide flexibility to the DS by demonstrating ranges of scientific capability, technical challenge, and cost, enabling an informed judgement on the right balance of all three factors.

*Why Now? Scientific and Technological Readiness*

The time for a mission like HabEx/LUVOIR is now. Recent advances in our scientific knowledge, largely through focused efforts by NASA, make it possible (for the first time) to design a robust mission concept that will allow for the detection and characterization of potentially habitable worlds. In particular, we now know that small, rocky planets are common. We also know that most stars are unlikely to have sufficiently massive exozodiacal belts to hinder the detection and characterization of Earthlike planets in the habitable zone. Finally, investments in starlight suppression technologies (coronagraphs and starshades), have lowered the risk of these mission concepts to an acceptable level. Furthermore, we have developed clear paths forward to move all technologies to TRL 6 by the beginning of Phase A.

Should we so choose, we can now start the development of a large strategic mission that will be a great successor to HST, and will not only answer some of the most fundamental questions in astrophysics and planetary science, but may also finally answer the question of whether or not there is life elsewhere in the Universe.